\documentclass[aps,pra,twocolumn]{revtex4}
\usepackage{graphicx}

\begin{document}

\title{BCS-BEC crossover and quantum phase transition for ${}^{6}$Li and ${}^{40}$K
atoms across Feshbach resonance}
\author{W. Yi and L.-M. Duan}

\address{FOCUS center and MCTP, Department of Physics, University of Michigan, Ann
Arbor, MI 48109}

\begin{abstract}
We systematically study the BCS-BEC crossover and the quantum
phase transition in ultracold ${}^{6}$Li and ${}^{40}$K atoms
across a wide Feshbach resonance. The background scattering
lengths for ${}^{6}$Li and ${}^{40}$K have opposite signs, which
lead to very different behaviors for these two types of atoms. For
$^{40}$K, both the two-body and the many-body calculations show
that the system always has two branches of solutions: one
corresponds to a deeply bound molecule state; and the other, the
one accessed by the current experiments, corresponds to a weakly
bound state with population always dominantly in the open channel.
For ${}^{6}$Li, there is only a unique solution with the standard
crossover from the weakly bound Cooper pairs to the deeply bound
molecules as one sweeps the magnetic field through the crossover
region. Because of this difference, for the experimentally
accessible state of $ ^{40} $K, there is a quantum phase
transition at zero temperature from the superfluid to the normal
fermi gas at the positive detuning of the magnetic field where the
s-wave scattering length passes its zero point. For $^{6}$Li,
however, the system changes continuously across the zero point of
the scattering length. For both types of atoms, we also give
detailed comparison between the results from the two-channel and
the single-channel model over the whole region of the magnetic
field detuning.
 \pacs{03.75.Ss, 05.30.Fk,34.50.-s}
\end{abstract}

\maketitle

\section{Introduction}

Ever since the experimental observation of the condensation of fermion pairs
in dilute gases of ultracold fermionic atoms, the study of the BCS-BEC
crossover in ultracold fermionic gases has attracted much attention \cite
{1,2}. The crossover in ultracold fermion gases is facilitated
experimentally by the Feshbach resonance, where the inter-atomic interaction
can be tuned over a large range by varying the external magnetic field \cite
{2'}. Various experiments have been done recently in ultracold fermionic
gases in the vicinity of the Feshbach resonance to study the properties of
the BCS-BEC crossover \cite{1}.

In most experiments, either an equal mixture of ${}^{40}$K atoms in the $%
|F=9/2,m_{F}=-9/2\rangle $ and the $|F=9/2,m_{F}=-7/2\rangle $ hyperfine
states or an equal mixture of ${}^{6}$Li atoms in the $|F=1/2,m_{F}=-1/2%
\rangle $ and the $|F=1/2,m_{F}=1/2\rangle $ states are prepared and
magnetically tuned across a wide Feshbach resonance ($B_{0}=202G$ for $%
{}^{40}$K and $B_{0}=834G$ for ${}^{6}$Li). Because of the nature
of the wide resonance, for both atom species, their properties
close to the resonance point are very similar \cite{3,4,5}.
However, they indeed have an important difference which is not
quite emphasized before: the background scattering lengths for
$^{40}$K and $^{6}$Li have opposite signs. This sign difference in
the background scattering has profound effects when the system is
outside the unitary region.

In this paper, we investigate the different properties of ${}^{40}$K and ${}^{6}$%
Li atoms arising from their distinction in the background
scattering. To describe the Feshbach resonance, we use the
standard two-channel model which is characterized by three
parameters determined from the scattering data: the resonance
width, the detuning, and the background scattering length \cite
{2,4,6}. The main differences between $^{40}$K and ${}^{6}$Li are
summarized as follows: First, for the $^{40}$K atoms, there always
exist two stable solutions to the many-body gap and number
equations; while for the $^{6}$Li atoms, there is only one
solution. The current experiments with $^{40}$K atoms probe only
one branch of the solution through the adiabatic sweep. For this
experimentally relevant branch, the closed channel population is
always limited to a small fraction, no matter how far the magnetic
field is detuned. This unusual property of $^{40}$K has been noted
before, in association with a somewhat different model Hamiltonian
which is characterized by five experimental parameters \cite{7,8}.
For future experiments with $^{40}$K, there might also be the
possibility to switch to the other branch of the solution through
a fast sweep, if the three-body collision does not turn out to be
a significant obstacle to the stability of such a state (note that
the three-body collision is not taken into account in the
two-channel Hamiltonian). Second, a probably more interesting
difference is that there exists a quantum phase transition at zero
temperature for the $^{40}$K atoms; while there is no such transition for $%
^{6}$Li. Note that here we limit ourselves to the case with equal
populations for the two spin components, where there is no phase
transition in the standard BCS-BEC crossover theory. Nevertheless,
for the experimentally accessible branch of solution of the
$^{40}$K atoms, due to the effect of the background scattering in
the two-channel model, a quantum phase transition from the
superfluid state to the normal Fermi gas occurs on the BCS side of
the resonance, basically around the point where the scattering
length goes across zero. For the ${}^{6}$Li atoms, however, when
the scattering length crosses the zero point, which is on the BEC
side, the change in the state of the system is completely
continuous.

In this work, we have also systematically compared the results
from the two-channel model and from the single-channel model over
the whole region of the magnetic field detuning (from the far
negative to the far positive). The standard single-channel model
is characterized by only a single parameter, the scattering length
at the corresponding field detuning. Our results extend some of
the comparisons in Refs. \cite{3} to the off-resonance region,
with particular emphasis on the effects arising from the
difference in the background scattering of the two atom species.
For easier comprehension of the properties of the many-body
solutions, we also present the solutions for the two-body states
with the same model Hamiltonian for comparison. The paper is
arranged as follows: In Sec. II, we present the model
Hamiltonians, and discuss their two-body states for ${}^{6}$Li and
${}^{40}$K atoms. Then, in Sec. III, we apply the two channel and
the single channel models respectively to study the many-body
properties of the ultracold atoms at zero temperature. The
properties calculated include the energy gap, the chemical
potential, and the closed channel population. In Sec. IV, we
analyze in detail the quantum phase transition for the ${}^{40}$K
atoms on the BCS side. Finally, in Sec. V, we discuss the finite
temperature behavior, and calculate the critical temperatures and
the pseudo-gaps
corresponding to the different branches of solutions for ${}^{6}$Li and $%
{}^{40}$K atoms. For this finite-temperature calculation, we follow the $%
G_{0}G$ method as reviewed in Ref. \cite{4} to include the pair
fluctuations. The results are summarized in Sec. VI.

\section{The Model Hamiltonians and their Two-body States}

The Feshbach resonance in ultracold atoms can be understood on the
grounds of two colliding fermions in different hyperfine states.
In dilute ultracold gases, the interaction can be described by the
s-wave atomic scattering. The magnetic field tunes the interaction
strength by Zeeman shifting the energy level of the quasi-bound
state (the Feshbach molecule state) relative to the continuum
threshold of free atoms. The physics is well described by the
following two-channel model \cite{2,6}
\begin{eqnarray}
H &=&\sum_{\sigma ,\mathbf{k}}\epsilon _{\mathbf{k}}a_{\mathbf{k},\sigma
}^{\dag }a_{\mathbf{k},\sigma }+\sum_{\mathbf{q}}(\gamma +\epsilon _{\mathbf{%
q}}/2)b_{\mathbf{q}}^{\dag }b_{\mathbf{q}}  \nonumber \\
&+&\left( U/\mathcal{V}\right) \sum_{\mathbf{k},\mathbf{k^{\prime }},\mathbf{%
q}}a_{\mathbf{k+q/2},\uparrow }^{\dag
}a_{-\mathbf{k-q/2},\downarrow }^{\dag
}a_{-\mathbf{k^{\prime }-q/2},\downarrow }a_{\mathbf{k^{\prime }+q/2}%
,\uparrow }\nonumber\\
&+& \left( g/\sqrt{\mathcal{V}}\right) \sum_{\mathbf{k},\mathbf{q}}\left( b_{%
\mathbf{q}}^{\dag }a_{-\mathbf{k+q/2},\downarrow
}a_{\mathbf{k-q/2},\uparrow }+h.c.\right),
\end{eqnarray}
where $a_{\mathbf{k},\sigma }^{\dag }$ and $b_{\mathbf{q}}^{\dag
}$ are the creation operators for the open channel fermions (the
atoms) and the closed channel bosons (the molecules),
respectively; $\epsilon _{\mathbf{k}}=\hbar ^{2}k^{2}/(2m)$ ($m$
is the atom mass); $\mathcal{V}$ is the quantization volume, and
$\sigma ={\uparrow ,\downarrow }$ denotes the different hyperfine
states of the atoms. This Hamiltonian is characterized by three
parameters: the bare atom-molecule coupling rate $g$, the bare
background atom scattering rate $U$, and the bare detuning $\gamma
.$ These three bare quantities are connected with the physical
ones $g_{p},U_{p},\gamma _{p}$
through the standard renormalization relations: $U=\Gamma U_{p}\,,$ $%
g=\Gamma g_{p}\,,$ $\gamma =\gamma _{p}-\Gamma g_{p}^{2}/U_{c}$, where $%
\Gamma \equiv \left( 1+U_{p}/U_{c}\right) ^{-1}$ and
$U_{c}^{-1}\equiv -\sum_{\mathbf{k}}\frac{1}{2\epsilon
_{\mathbf{k}}}$ \cite{4}. The physical quantities
$g_{p},U_{p},\gamma _{p}$ are determined from the scattering data
as $U_{p}=4\pi \hbar ^{2}a_{bg}/m$, $g_{p}^2=4\pi \hbar
^{2}a_{bg}\Delta B\mu _{co}/m$, and $\gamma _{p}=\mu
_{co}(B-B_{0})$ ($\mu _{co}$ is the difference in magnetic moments
between the two channels) \cite{9}, where we
have assumed that the s-wave scattering length near resonance has the form $%
a_{s}=a_{bg}\left( 1-\frac{\Delta B}{B-B_{0}}\right) $, with $a_{bg}$ as the
background scattering length, $\Delta B$ as the resonance width, and $B_{0}$
as the resonance point.

If the population in the closed channel is negligible, one can adiabatically
eliminate the molecular modes $b_{\mathbf{q}}$ in the Hamiltonian (1), and
arrive at the following simplified single-channel model
\begin{eqnarray}
&H&=\sum_{\sigma ,\mathbf{k}}\epsilon
_{\mathbf{k}}a_{\mathbf{k},\sigma
}^{\dag }a_{\mathbf{k},\sigma }\\&+&\left( U^{eff}/\mathcal{V}\right) \sum_{%
\mathbf{k},\mathbf{k^{\prime }},\mathbf{q}}a_{\mathbf{k+q/2},\uparrow
}^{\dag }a_{-\mathbf{k-q/2},\downarrow }^{\dag }a_{-\mathbf{k^{\prime }-q/2}%
,\downarrow }a_{\mathbf{k^{\prime }+q/2},\uparrow },\nonumber
\end{eqnarray}
where $U^{eff}$ is connected with the physical one $U_{p}^{eff}$
by the renormalization relation $U^{eff}=U_{p}^{eff}\left(
1+U_{p}^{eff}/U_{c}\right) ^{-1}$, and $U_{p}^{eff}$ is determined
solely by the scattering length $a_{s}$ as $U_{p}^{eff}=4\pi \hbar
^{2}a_{s}/m$ at the corresponding magnetic field detuning. So, a
single-channel model does not explicitly depend on the resonance
width or the background scattering rate, it only depends on their
combination through the expression of $a_{s}$. In this sense, a
single-channel model is simpler and more universal. However, given
a new configuration, we typically do not know whether the closed
channel population is negligible prior to the calculation,
therefore the validity of the single-channel model is not
self-evident. We will have some systematic comparison of the
results from the two-channel and the single-channel calculations
for $^{40}$K and $^{6}$Li over the whole region of the magnetic
field detuning. Similar comparison has been done before in the
unitary region \cite{3,4}, where the background scattering can be
neglected.

To have some intuitive understandings of the Hamiltonian above, we
first look at its bound states for two particles. For a two
channel model, the eigenstate of the Hamiltonian in the case of
two particles can be written as the superposition of the molecular
bosons in the closed channel and the fermion pairs in the open
channel:
\begin{equation}
\left| \Psi \right\rangle =\sum_{\mathbf{k}}c_{\mathbf{k}}a_{\mathbf{k}%
,\uparrow }^{\dag }a_{-\mathbf{k},\downarrow }^{\dag }+c_{b}b_{0}^{\dag
}\left| 0\right\rangle ,
\end{equation}
where $c_{\mathbf{k}}$ and $c_{b}$ are the superposition coefficients, and
we have assumed that the center-of-mass momentum $\mathbf{q=0}$ as it is
decoupled from the relative momentum in the two-body case. The Schrodinger's
equation $H\left| \Psi \right\rangle =E\left| \Psi \right\rangle $ then
gives:
\begin{eqnarray}
\frac{1}{U_{p}-\frac{g_{p}^{2}}{\gamma _{p}-E}}&=&-\frac{1}{\mathcal{V}}\sum_{%
\mathbf{k}}\left( \frac{1}{2\epsilon _{\mathbf{k}}-E}-\frac{1}{2\epsilon _{%
\mathbf{k}}}\right) \nonumber\\
&=&2^{-\frac{7}{2}}\pi^{-1}\left(\frac{\hbar^2}{2m}\right)^{-\frac{3}{2}}\sqrt{-E},
\end{eqnarray}
where we have used the renormalization relation $1/\left[ U_{p}-g_{p}^{2}/%
\gamma _{p}\right] =1/\left[ U-g^{2}/\gamma \right] +(1/\mathcal{V})\sum_{%
\mathbf{k}}\left[ 1/\left( 2\epsilon _{\mathbf{k}}\right) \right] $.

We solve the bound state energy levels of ${}^{6}$Li and ${}^{40}$K,
respectively. The parameters for the ${}^{6}$Li atoms are $a_{bg}=-1405a_{0}$%
, $\Delta B=-300G$ and $\mu _{co}\sim 2\mu _{B}$, where $a_{0}$ is
the Bohr
radius and $\mu _{B}$ is the Bohr magnetic moment. The parameters for the $%
{}^{40}$K atoms are $a_{bg}=174a_{0}$, $\Delta B_{0}=7.8G$, and
$\mu _{co}\sim 1.68\mu _{B}$. The results of our calculation are
plotted in Fig. 1. The outstanding feature of the figures is that
there are two branches of bound state for ${}^{40}$K (Fig. 1(a)).
One branch is a deep-bound molecule state which never crosses the
continuum threshold (zero energy in the plots). In this state, the
closed channel fraction ($\left| c_{b}\right| ^{2}$) becomes
dominant as one approaches the negative detuning. The other branch
(the upper one) is a
weakly bound pair state which crosses the continuum threshold at $%
\gamma =0$. In this state, the closed channel fraction is always
limited to a small fraction ($<6\%$), no matter how much the
magnetic field is detuned. In experiments, one approaches the
bound state by adiabatically tuning the magnetic field, starting
from the atom continuum. Therefore the branch of state with weakly
bound pairs is the one accessed in the current experiments. If one
wants to go to the deeply bound molecule state, one either needs a
fast magnetic field sweep with a rate larger then the energy
splitting between the two bound states, or use some rf pulses to
couple these two branches. For the ${}^{6}$Li atoms, in contrast,
there is only one branch of solution, which merges into the continuum at $%
\gamma =0$ and approaches the deep-bound molecule state as one
increases the negative field detuning. This difference between the
two atom species comes from the different natures of their
background scattering interactions. For $^{40}$K with a positive
background scattering length, as has been discussed in Refs.
\cite{7,8,10} (the Hamiltonians therein are characterized by five
parameters and are somewhat different from the model here), there
is a weakly bound state in the open collision channel. The avoided
level crossing between this weakly bound state and the Feshbach
molecule level gives the two branches of solutions. For $^{6}$Li
with a negative background scattering length, there is no bound
state in the open channel, and thus there is only a single branch
of solution.
\begin{figure}[tbp]
\includegraphics{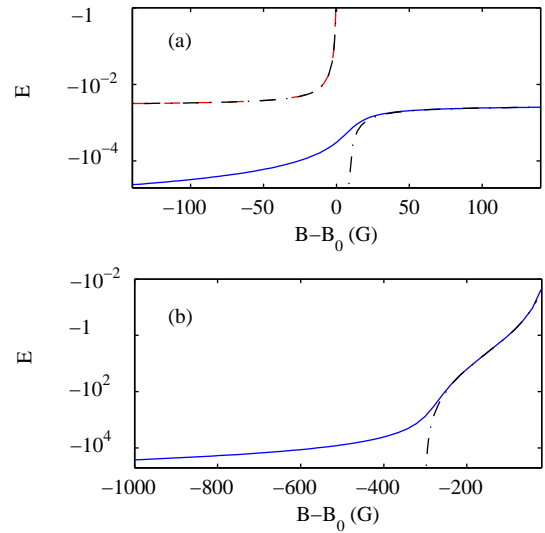}
\caption[Fig.1 ]{The bound state energy levels for two colliding atoms of (a)%
${}^{40}$K, (b)${}^{6}$Li. The solid curves are the deeply bound
states, the dashed curve is the weakly bound state in ${}^{40}$K,
and the dash-dotted curves are the results from single channel
Hamiltonian. To compare with the manybody calculations that
follow, the unit of energy is taken to be $E_{F}=\hbar
^{2}(3\protect\pi ^{2}n)^{2/3}/2m$, where the total
particle densities ($n$) are chosen to be typical values in the experiments (%
$n\sim 1.8\times 10^{13}cm^{-3}$ for ${}^{40}$K, and $n\sim
2.9\times 10^{13}cm^{-3}$ for ${}^{6}$Li \cite{1} ). }
\end{figure}

For comparison, we have also calculated the two-body bound states
using the single channel Hamiltonian (2). In the single-channel
approach, one simply replaces $U_{p}-\frac{g_{p}^{2}}{\gamma
_{p}-E}$ in Eq. (4) with $U_{p}^{eff} $, so the left side of Eq.
(4) becomes energy independent. Equation (4) then gives the well
known expression of the bound energy $E=-\hbar ^{2}/\left(
ma_{s}^{2}\right) $ in the region $a_{s}\geq 0$ for the
single-channel mode. The results of the calculation are shown as
the dash-dotted curves in Fig. 1. The single channel calculation
gives almost the same bound state energies for the weakly bound
state of ${}^{40}$K atoms (it neglects the other branch of
solution); while it deviates from the results of the two channel
calculation as the atomic scattering length $a_{s}$ approaches its
zero point (corresponding to $B-B_{0}=7.8G$ for ${}^{40}$K, and
$B-B_{0}=-300G$ for ${}^{6}$Li).

\section{The Many-body States}

To find out the many-body state at zero temperature, one needs to minimize
the energy corresponding to the Hamiltonian (1) under the constraint of
total particle number conservation $N\equiv \sum_{\sigma ,\mathbf{k}%
}a_{\mathbf{k},\sigma }^{\dag }a_{\mathbf{k},\sigma }+2\sum_{\mathbf{q}}b_{%
\mathbf{q}}^{\dag }b_{\mathbf{q}}$. One therefore minimizes
$\Omega =\left\langle H-\mu N\right\rangle $, where $\mu $ is the
Lagrange multiplier with the physical meaning of the chemical
potential. We follow the standard BCS-BEC crossover theory
\cite{11}, which assumes a mean
field for the molecule and the pair operators, with $\langle b_{\mathbf{q}%
}\rangle =\langle b_{0}\rangle \delta _{\mathbf{q0}}=-\left( g/\sqrt{%
\mathcal{V}}\right) \sum_{\mathbf{k}}\left\langle a_{-\mathbf{k},\downarrow
}a_{\mathbf{k},\uparrow }\right\rangle /\left( \gamma -2\mu \right) $, where
the second equality comes from the Heisenberg equation for the operator $b_{%
\mathbf{q}}$. Under such an assumption, one can find out the explicit
expression of the energy $\Omega $, whose extrema conditions $\frac{%
\partial \Omega }{\partial \langle b_{0}\rangle }=\frac{\partial \Omega }{%
\partial \mu }=0$ yield the standard gap equation and the number equation:
\begin{eqnarray}
\frac{1}{U_{p}-\frac{g_{p}^{2}}{\gamma _{p}-2\mu }} &=&-\frac{1}{\mathcal{V}}%
\sum_{\mathbf{k}}(\frac{1}{2E_{\mathbf{k}}}-\frac{1}{2\epsilon _{\mathbf{k}}}%
), \\
n_{tot} &=&2n_{b}+\frac{1}{\mathcal{V}}\sum_{\mathbf{k}}(1-\frac{\epsilon _{%
\mathbf{k}}-\mu }{E_{\mathbf{k}}}),
\end{eqnarray}
where the quasi-particle excitation energy $E_{\mathbf{k}}=\sqrt{(\epsilon _{%
\mathbf{k}}-\mu )^{2}+\left| \Delta \right| ^{2}}$, and the gap $\Delta
=g\langle b_{0}\rangle /\sqrt{\mathcal{V}}+\left( U/\mathcal{V}\right) \sum_{%
\mathbf{k}}\langle a_{-\mathbf{k},\downarrow }a_{\mathbf{k},\uparrow
}\rangle =z_{b}\langle b_{0}\rangle /\sqrt{\mathcal{V}}$ ($z_{b}\equiv
g-U\left( \gamma -2\mu \right) /g=g_{p}-U_{p}\left( \gamma _{p}-2\mu \right)
/g_{p}$). In Eq. (6), $n_{tot}$ denotes the total particle density, and $%
n_{b}=\left| \left\langle b_{0}\right\rangle \right| ^{2}/\mathcal{V}$ is
the density of the closed channel molecules.

With Eqs.(5) (6), we can solve for the gap, the chemical potential
and the closed channel fraction at different magnetic field
detunings. The results of the calculations for ${}^{40}$K and for
${}^{6}$Li are plotted in Fig. 2 and Fig. 3, respectively. For
${}^{40}$K, similar to the two-body case, Eqs. (5) (6) support two
branches of solutions. The properties of the first branch of
solutions is shown in Fig. 2 (a-c). Compared with the results in
the previous section, it is easy to see that this solution
corresponds to the deep-bound state in the two-body energy
spectrum. The closed channel fraction increases from $0$ in the
BCS limit to $1$ in the BEC limit, and
the maximum in the gap appears on the BCS side of the resonance, near the $%
a_{s}=0$ point. As the chemical potential always remains negative for this
branch, the system is essentially BEC-like at all magnetic field detunings.
The properties of the second branch of solution is shown in Fig. 2 (d-f),
where the closed channel fraction is non-monotonic, peaks around $%
B-B_{0}\sim -5.2G$ with a fraction $\sim5.1\%$, in rough agreement
with the calculations in Refs. \cite{7,8}, where a different model
is used. The chemical potential in this branch crosses zero on the
BEC side of the resonance, demonstrating the existence of a
BCS-BEC crossover for this branch. The most intriguing feature of
the second branch is that there is a quantum phase transition on
the BCS side of the resonance, beyond which the superfluid
disappears, and the system phase-transits into a normal fermi gas.
We will discuss this phase transition in more details in the next
section. The properties of ${}^{6}$Li atoms are shown in Fig. 3.
The closed channel fraction remains small throughout the crossover
region, and only becomes appreciable deep in the BEC region. The
peak in the gap occurs again near the $a_{s}=0$ point (now on the
BEC side). The chemical potential crosses zero around
$B-B_0\sim-85G$, denoting a BCS-BEC crossover.

\begin{figure}[tbp]
\includegraphics{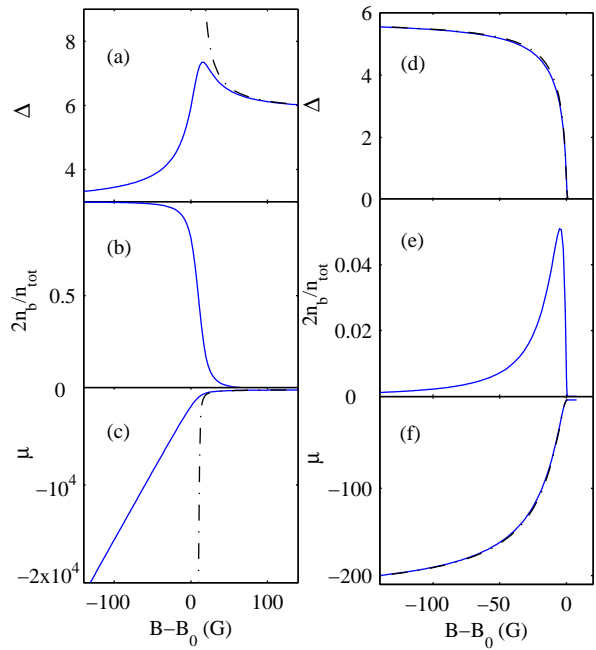}
\caption[Fig.2 ]{Results from the manybody calculation for
${}^{40}$K atoms at zero temperature. (a-c) The gap, the closed
channel fraction, and the chemical potential of the solution which
corresponds to the deeply bound state in the twobody case; (d-f)
The gap, the closed channel fraction, and the chemical potential
of the solution which corresponds to the weakly bound state in the
twobody case. The solid curves are the results using two-channel
Hamiltonian. The dash-dotted curves are the results using single
channel Hamiltonian. The unit of energy is the same as defined in
the caption of Fig. 1.}
\end{figure}

For comparison, we have calculated the gap and the chemical potential using
the single channel Hamiltonian. For the single-channel approach, one just
replaces $U_{p}-\frac{g_{p}^{2}}{\gamma _{p}-2\mu }$ with $U_{p}^{eff}$ in
the gap equation (5) and drops the closed channel population contribution $%
2n_{b}$ in the number equation (6). The results are shown in Fig.
2 and Fig. 3 in dash-dotted curves. The results from the single
channel calculation should coincide with the two channel
calculation provided that the closed channel population is small.
From Fig. 2 and Fig. 3, it is obvious that, except for the
deep-bound state solution of ${}^{40}$K, the single channel
Hamiltonian reproduces the results of the two channel Hamiltonian for $%
{}^{40}$K and ${}^{6}$Li for most of the crossover region. Notably, the
single channel description is a fairly good approximation for almost all
magnetic field detunings for the weakly bound state solution of ${}^{40}$K,
as the closed channel fraction in this branch remains small. In both cases,
the single channel description breaks down near the point where $a_{s}=0$.
\begin{figure}[tbp]
\includegraphics{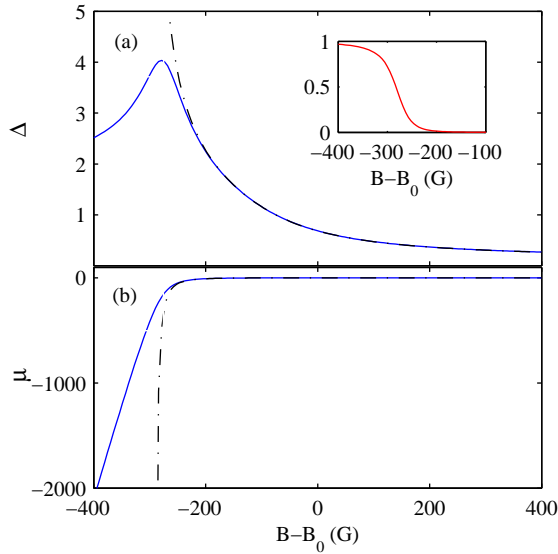}
\caption[Fig.3 ]{Results from the manybody calculation for
${}^{6}$Li atoms at zero temperature. (a)The gap and the closed
channel fraction (inset); (b) The chemical potential. The solid
curves are the results using two-channel Hamiltonian. The
dash-dotted curves are the results using single channel
Hamiltonian. }
\end{figure}

\section{Quantum Phase Transition for ${}^{40}$K atoms}

To see more clearly the quantum phase transition, we plot the gap and the
closed channel fraction from the two channel calculation near the phase
transition point on a logarithmic scale in Fig. 4. Both the gap and the
closed channel fraction show a precipitous drop at the magnetic detuning $%
B-B_{0}\sim 7.8G$, the point where $a_{s}=0$. Beyond that point,
the gap equation and the number equation do not have solutions
(for the upper branch). This is a clear signature of a quantum
phase transition. The ground state of the ${}^{40}$K atoms changes
from a BCS superfluid state into a normal state with the gap
$\Delta =0$ through a second order phase transition. In contrast,
for ${}^{6}$Li atoms (Fig. 3), there is just a smooth crossover
for both the gap and the closed channel population at the point
where $a_{s}=0$ ($B-B_{0}=-300G$).

The origin of the phase transition comes from the repulsive background
inter-atomic interaction between the ${}^{40}$K atoms in the two different
hyperfine states. The repulsive interaction does not support a superfluid
state which can be adiabatically connected to the BCS superfluid in the region $%
a_{s}<0$ (although the system indeed has a superfluid solution
corresponding to the deep-bound state, it is not adiabatically
connected). So, across the point $a_{s}=0$, there must be a phase
transition, and the atoms transit into a normal phase when
$a_{s}>0$. If one starts from this
normal Fermi gas, as the magnetic field decreases towards the resonance ($%
B_{0}=202G$), the atoms will first undergo a second order phase
transition and become a BCS superfluid at $B-B_{0}\sim 7.8G$. If
the field continues to decrease, the chemical potential will
eventually become negative at $B-B_{0}\sim -0.34G$, and the atoms
enter the BEC region with a smooth crossover.
\begin{figure}[tbp]
\includegraphics{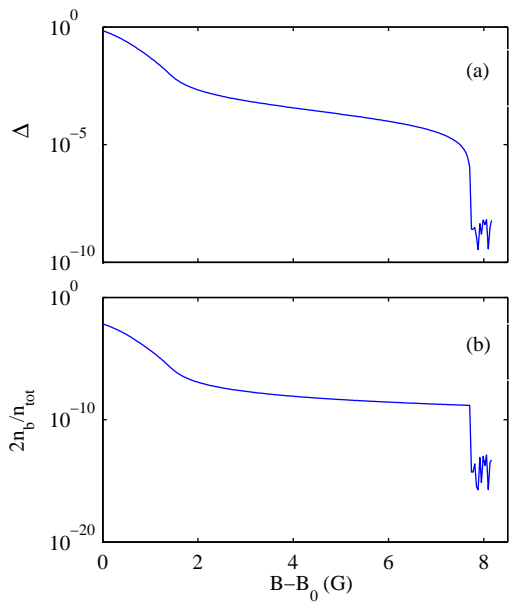}
\caption[Fig.4 ]{The enlarged (a) gap and (b) closed channel
fraction as a function of magnetic field detuning for ${}^{40}$K
atoms near the phase transition ($B-B_{0}\sim 7.8G$)
(corresponding to the weakly bound branch of solution).}
\end{figure}

\section{The Critical Temperatures for Different Bound States}

To characterize the crossover properties of different bound states for
different atom species, we calculate the critical temperature at which all
the pairs become non-condensed. At finite temperature $T$, instead of the
energy minimization, one needs to minimize the thermodynamical potential $%
\Omega =-T\ln [$tr$\left( e^{-H/T}\right) ]$. If one takes the
mean-field
approach outlined in Sec. II, from the extrema conditions $%
\frac{\partial \Omega }{\partial \langle b_{0}\rangle }=\frac{\partial
\Omega }{\partial \mu }=0$, one gets the following finite-temperature
version of the gap equation and the number equation \cite{12,2}
\begin{eqnarray}
\frac{1}{U_{p}-\frac{g_{p}^{2}}{\gamma _{p}-2\mu }} =-\frac{1}{\mathcal{V}}%
\sum_{\mathbf{k}}\left(\frac{1-2f(E_{\mathbf{k}})}{2E_{\mathbf{k}}}-\frac{1}{2\epsilon _{\mathbf{k}}}%
\right),\\
n_{tot}
=2n_{b}+\frac{1}{\mathcal{V}}\sum_{\mathbf{k}}\left(1-\frac{\epsilon
_{ \mathbf{k}}-\mu }{E_{\mathbf{k}}}+2\frac{\epsilon _{
\mathbf{k}}-\mu }{E_{\mathbf{k}}}f(E_{\mathbf{k}})\right),
\end{eqnarray}
where $f(x)=1/(e^{x/k_BT}+1)$ is the fermi distribution, and $k_B$
is the Boltzmann constant.

This simple mean-field approach however, is not adequate for the
calculation of the critical temperature except in the case of the
BCS limit. The reason is that at finite $T$, due to the thermal
fluctuations, there exist non-condensed pairs (superposition of
fermion pairs and non-condensed bosons), in addition to the
condensed fermion pairs and the condensed molecular bosons
\cite{4,11}. The non-condensed pairs also contribute to the gap
for the fermionic excitations. Accommodating this contribution,
the gap is replaced by the total gap $\left| \Delta \right|
^{2}=\left| \Delta _{s}\right| ^{2}+\left| \Delta _{pg}\right|
^{2}$, where $\Delta _{s}=z_{b}\langle b_{0}\rangle
/\sqrt{\mathcal{V}}$ is the same as the gap defined in the
previous section (the superfluid order parameter), and the
pseudo-gap $\Delta _{pg}$ comes from the contributions of the
non-condensed pairs \cite{4,11}. With this simple replacement, the
above gap equation and the number equation from the mean-field
approach remain valid. One just needs to interpret $\Delta $ as
the total gap, and thus the emergence of $\Delta $ does not
correspond to the critical transition between the normal and the
superfluid phase. Instead, the gap shows up at a temperature
$T^{\ast }$, below which the non-condensed pairs are formed. As
one further decreases the temperature, these pre-formed pairs
finally condense and become a superfluid at the critical
temperature $T_{c}$ with $T_{c}<T^{\ast
}$. The gap at the critical temperature $T_c$ solely comes from the pseudo-gap $%
\Delta _{pg}$, contributed by the non-condensed pairs (as $\Delta
_{s}=0$ at that point). To calculate this phase transition
temperature $T_{c}$, one needs to breakup the total gap $\Delta $
into the superfluid order parameter $\Delta _{s}$ and the
pseudo-gap $\Delta _{pg}$, which requires knowledge of the
dispersion relation of the pair excitations. As shown in Ref.
\cite{4}, the dispersion relation for the pair excitations is
still quadratic with a self-consistently determined effective mass
$M^{\ast }$. With that formalism, we calculate the critical
temperature $T_{c}$ and the pseudo-gap at $T_{c}$ for different
branches of states in $^{40}$K and ${}^{6}$Li over the whole
region of the magnetic field detuning.

The results of the calculation are plotted in Fig. 5 for
${}^{40}$K atoms, and in Fig. 6 for ${}^{6}$Li atoms. Fig. 5 (a)
(b) correspond to the deeply bound state, and Fig. 5 (c) (d)
correspond to the case with weakly bound state. It is clear from
the $T_{c}$ calculation that the BCS-BEC crossover picture is
evident only for the second branch of the solution (Fig. 5 (c)
(d)). For this branch, the critical temperature becomes
exponentially small on the BCS side and almost constant on the BEC
side as one expects. For the other branch of the solution
corresponding to the deep-bound state, the critical temperature
only varies in a very small range, and converges to a constant
value on either side of the resonance. One can also see that
except in the BCS limit, the pseudo-gap is still significant at
the critical temperature.
\begin{figure}[tbp]
\includegraphics{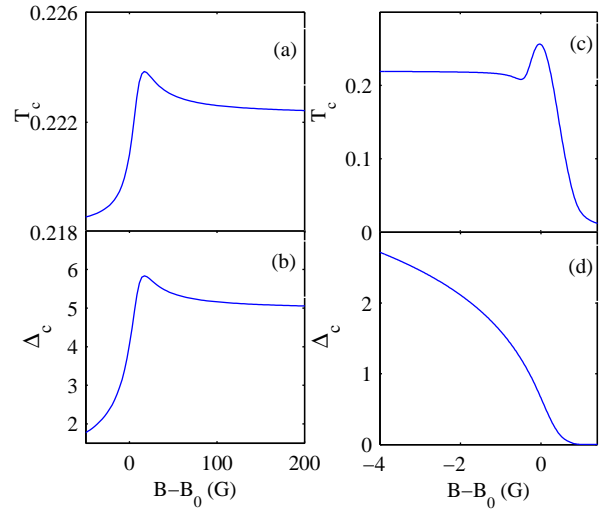}
\caption[Fig.5 ]{The critical temperature $T_{c}$ and the
pseudogap at the critical temperature $\Delta _{c}$ as a function
of magnetic field detuning for the two solutions in ${}^{40}$K
atoms ((a)(b) correspond to the deeply bound branch, while (c) (d)
correspond to the weakly bound branch). $T_{c}$ is in the unit of $T_{F}$%
, the Fermi temperature associated with the Fermi energy defined in the
caption of Fig. 1.}
\end{figure}
\begin{figure}[tbp]
\includegraphics{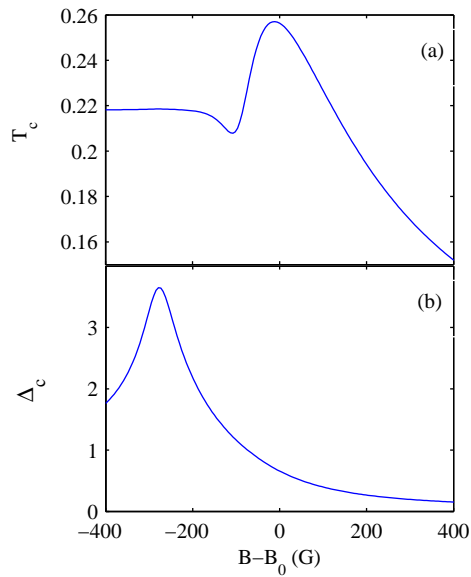}
\caption[Fig.6]{The critical temperature $T_{c}$ and the pseudogap
at the critical temperature $\Delta _{c}$ as a function of
magnetic field detuning in ${}^{6}$Li atoms. $T_{c}$ is in the
unit of $T_{F}$, the Fermi temperature associated with the Fermi
energy defined in the caption of Fig. 1.}
\end{figure}

\section{Summary}

We have systematically studied the BCS-BEC crossover and the
quantum phase transition in ultracold ${}^{6}$Li and ${}^{40}$K
atoms across a wide Feshbach resonance. As the background
scattering lengths have opposite signs for these two types of
atoms, their properties are quite different outside of the near
resonance region. Both the two-body and the many-body calculations
show that for $^{40}$K, there always exist two branches of
solutions, corresponding respectively to the deep-bound state and
the weakly bound states. The latter one with weakly bound pairs is
the branch that is accessed by the current experiments through
adiabatic sweep. For this branch of solution, there is a quantum
phase transition on the BCS side of the resonance, where the atoms
phase-transit from the superfluid state to the normal fermi gas as
the scattering length crosses zero. This kind of phase transition
is absent in the ${}^{6}$Li atoms, where the change in the state
of the system is completely continuous when the scattering length
goes across its zero point.

The calculations here emphasize the importance of the background
scattering length outside of the near-resonance region, and the
different properties of the $^{6}$Li and ${}^{40}$K atoms. In this
sense,  $^{6}$Li and ${}^{40}$K are good representatives of two
kinds of atoms which have different background scattering
properties. The calculations here may also have some interesting
experimental indications. For instance, one can verify the quantum
phase transition for the $^{40}$K atoms on the BCS\ side of the
resonance, and the continuous crossover for the $^{6}$Li atoms on
the BEC side of the resonance, although in both cases the
scattering length crosses its zero point in a similar way. For the
$^{40}$K atoms, there might also be the possibility to access the
deeply bound branch of solution in future experiments through
either a fast magnetic field sweep or some rf transitions.

This work was supported by the NSF awards (0431476), the ARDA
under ARO contracts, and the A. P. Sloan Fellowship. We thank
Qijin Chen for helpful discussions.

\end{document}